\documentstyle[a4wide,epsf,11pt,titlepage]{article}

\newcounter{nref}
\newcommand{\bbib}{%
  \renewcommand{\refname}{\large\bf References}%
  \begin{thebibliography}{99}%
  \setcounter{enumiv}{\thenref}}
\newcommand{\ebib}{%
  \end{thebibliography}%
  \setcounter{nref}{\arabic{enumiv}}}
\newcommand{\head}[3]{%
  \setcounter{nref}{0}%
  \thispagestyle{empty}%
  \section*{\LARGE\bf #1}%
  \stepcounter{section}%
  \addcontentsline{toc}{section}{#1}%
  \large\itshape%
  #2\\\vspace{0.1pt}\\%
  #3%
  \normalsize\upshape%
  \bigskip}

\begin{document}


\head{Superstatistical turbulence models}
     {Christian Beck}
     {School of Mathematical Sciences, Queen Mary, University
of London, Mile End Road, London E1 4NS, UK}






\subsection*{Abstract}
Recently there has been some progress in modeling the statistical
properties of turbulent flows using simple superstatistical
models. Here we briefly review the concept of superstatistics in
turbulence. In particular, we discuss a superstatistical
extension of the Sawford model and compare with experimental data.

\subsection*{$\,$}


Turbulence is a spatio-temporal chaotic dynamics generated by the
Navier-Stokes equation
\begin{equation}
\dot{\vec{v}}=-(\vec{v}\nabla)\vec{v}+ \nu \Delta
\vec{v} +\vec{F} .
\end{equation}
In the past 5 years there has been some experimental progress in
Lagrangian turbulence measurements, i.e.\ tracking single tracer
particles in the turbulent flow. Due to the measurements of the
Bodenschatz \cite{beck.boden1,beck.boden2,beck.reynolds-recent} and Pinton
groups \cite{beck.pinton1,beck.pinton2} we now have a better view of what
the statistics of a single test particle in a turbulent flow looks
like. The recent measurements have shown that the acceleration
$\vec{a}$ as well as velocity difference
$\vec{u}=\vec{v}(t+\tau)-\vec{v}(t)$ on short time scales $\tau$
exhibits strongly non-Gaussian behavior. This is true for both,
single components as well as the absolute value of $\vec{a}$ and
$\vec{u}$. Moreover, there are correlations between the various
components of $\vec{a}$, as well as between velocity and
acceleration. The corresponding joint probabilities do not
factorize. Finally, the correlation functions of the absolute
value $|\vec{a}|$ and $|\vec{u}|$ decay rather slowly.

How can we understand all this by simple stochastic models?
There is a
recent class of models that are pretty successful in explaining
all these statistical properties of Lagrangian turbulence (as well
as of other turbulent systems, such as Eulerian turbulence
\cite{beck.BLS, beck.beck-physica-d, beck.jung-swinney}, atmospheric turbulence
\cite{beck.rapisarda, beck.rap2, beck.peinke} and defect turbulence 
\cite{beck.daniels}).
These are turbulence models based on superstatistics
\cite{beck.beck-cohen}. 
Superstatistics is a concept from
nonequilibrium statistical mechanics, in short it means a
`statistics of statistics', one given by ordinary Boltzmann factors and
another one given by fluctuations of an intensive parameter,
e.g.\ the inverse temperature, or the energy dissipation,
or a local variance. While the idea of
fluctuating intensive parameters is certainly not new, it is the
application to spatio-temporally chaotic systems such as
turbulent flow that makes the concept interesting. The first 
turbulence model of this kind 
was introduced in \cite{beck.prl}, in the meantime
the idea has been further refined and extended
\cite{beck.reynolds,beck.beck03,beck.reynolds-recent,beck.jung-swinney}. 
The basic idea is to generate a
superposition of two statistics, in short a `superstatistics', by
stochastic differential equations whose parameters fluctuate on a
relatively large spatio-temporal scale. In Lagrangian turbulence,
this large time scale can be understood by the fact that the
particle is trapped in vortex tubes for quite a while
\cite{beck.reynolds-recent}. Superstatistical turbulence models
reproduce all the experimental data quite well. An example is
shown in Fig.~1.
\begin{figure}
 \centerline{\epsfxsize=0.6\textwidth\epsffile{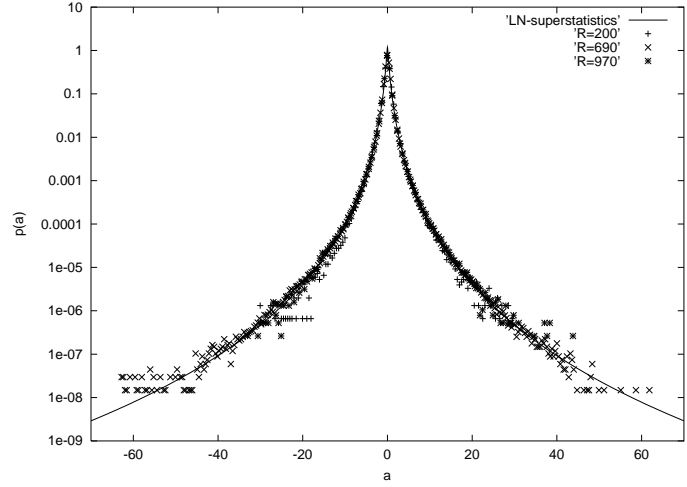}}
\caption{Probability
density of an acceleration component of a tracer particle as
measured by Bodenschatz et al. \cite{beck.boden1, beck.boden2}. The solid 
line is a theoretical prediction based on lognormal
superstatistics ($s^2=3$) \cite{beck.beck03}.}
\end{figure}
The theoretical prediction which fits the data perfectly is given
by
\begin{equation}
p(a)=\frac{1}{2\pi s} \int_0^\infty d\beta \beta^{-1/2} \exp
\left\{ -\frac{(\log (\beta/\mu))^2}{2s^2} \right\}
e^{-\frac{1}{2}\beta a^2} \label{beck.paln}
\end{equation}
with $\mu =e^{\frac{1}{2}s^2}$ and only one fitting parameter,
$s^2=3.0$. A similar formula as eq.~(\ref{beck.paln}) was already
considered in \cite{beck.castaing}, though without a dynamical
interpretation in terms of a stochastic differential equation
with fluctuating parameters.

The key ingredient of superstatistical models is to start from a
known model generating Gaussian behaviour, and extend it to a
superstatistical version exhibiting `fat tails'. In general, in
these types of models one has for some dynamical variable $a$ the
stationary long-term density
\begin{equation}
p(a)=\int_0^\infty \sqrt{\frac{\beta}{2\pi}}f(\beta)
e^{-\frac{1}{2}\beta a^2}d\beta , \label{beck.pa}
\end{equation}
where $f(\beta )$ is some suitable probability density of a
fluctuating parameter $\beta$. The function
$f(\beta)$ fixes the type of superstatistics under consideration.
In particular, it is responsible for the shape of the tails
\cite{beck.touchbeck}. Note the mixing of two statistics, that
of $a$ and that of $\beta$.

In Lagrangian turbulence, one may first start from a Gaussian
turbulence model, the Sawford model \cite{beck.sawford, beck.pope}. This
model considers the joint stochastic process $(a(t),v(t),x(t))$ of
an arbitrary component of acceleration, velocity and position
of a Lagrangian test particle, and assumes that they
obey the stochastic differential equation
\begin{eqnarray}
\dot{a}& =&-(T_L^{-1}+t_\eta^{-1})a-T_L^{-1}t_\eta^{-1} v\nonumber
\\ &\,&
+\sqrt{2\sigma_v^2(T_L^{-1}+t_\eta^{-1})T_L^{-1}t_\eta^{-1}}\;
L(t)
\\ \dot{v} &=&a \\ \dot{x} &=&v,
\end{eqnarray}

\indent $L(t)$: Gaussian white noise

$T_L$ and $t_{\eta}$:  two time scales, with $T_L
>>t_\eta$,

$T_L=2\sigma_v^2/(C_0 \bar{\epsilon})$

$t_\eta = 2a_0\nu^{1/2}/(C_0\bar{\epsilon}^{1/2})$

$\bar{\epsilon}$: average energy dissipation

$C_0, a_0$: Lagrangian structure function constants

$\sigma_v^2$ variance of the velocity distribution

$ R_\lambda = \sqrt{15}\sigma_v^2/\sqrt{\nu \bar{\epsilon}} $
Taylor scale Reynolds number.

\noindent For our purposes it is sufficient to consider the limit $T_L \to
\infty$, which is a good approximation for large Reynolds
numbers. In that limit the Sawford model reduces to just a linear
Langevin equation

\begin{equation}
\dot{a}=-\gamma a +\sigma L(t) \label{beck.la}
\end{equation}
with
\begin{eqnarray}
\gamma &=&\frac{C_0}{2a_0} \nu^{-1/2} \bar{\epsilon}^{1/2}\label{beck.gamma} \\
\sigma &=& \frac{C_0^{3/2}}{2a_0} \nu^{-1/2} \bar{\epsilon}.
\label{beck.sigma}
\end{eqnarray}
Note that this is a Langevin equation for the acceleration, not for
the velocity, in marked contrast to ordinary Brownian motion.

Unfortunately, the Sawford model predicts Gaussian stationary
distributions for $a$, and is thus at variance with the
recent measurements. So how can we save this model?

As said before, the idea is to generalize the Sawford model with
constant parameters to a superstatistical version. 
To construct a superstatistical extension of
Sawford model, we replace in the above equations the constant
energy dissipation $\bar{\epsilon}$ by a fluctuating one. One
formally defines a variance parameter \cite{beck.beck03}
\begin{equation}
\beta := \frac{2\gamma}{\sigma^2} = \frac{4a_0}{C_0^2} \nu^{1/2}
\frac{1}{\epsilon^{3/2}}, \label{beck.26}
\end{equation}
where $\epsilon$ fluctuates.
Now, if $\beta$ varies on a large spatio-temporal scale, and is
distributed with the distribution $f(\beta)$, one ends up with
eq.~(\ref{beck.pa}) describing the long-term marginal distribution of
the superstatistical dynamics (\ref{beck.la}). This is basically the
type of model introduced in \cite{beck.prl}, there with $f(\beta)$
chosen to be a $\chi^2$-distribution. Models based on
$\chi^2$-superstatistics yield good results for atmospheric
turbulence \cite{beck.rapisarda, beck.rap2}, and ultimately lead to Tsallis
statistics \cite{beck.tsa1}. On the other hand, for laboratory
turbulence experiments one usually obtains better agreement with
experimental data if $f(\beta)$ is a lognormal
distribution. In view of eq.~(\ref{beck.26}) this is clearly motivated
by Kolmogorov's ideas of a lognormally distributed $\epsilon$
\cite{beck.k62}.


Superstatistical models are not restricted to Lagrangian
turbulence but can be also formulated for Eulerian turbulence
\cite{beck.beck-physica-d, beck.prl}. Fig.~2 shows that also here one
obtains excellent agreement with experimental data: Probability
densities $p(u)$ of longitudinal velocity differences $u$ are
well fitted by lognormal superstatistics on all scales. The
parameter $s^2$ varies with the scale. In fact, not only the
distribution $p(u)$ but also the distribution $f(\beta)$ can be
directly measured in experiments \cite{beck.jung-swinney}, and the
two can be consistently connected via the superstatistics
formalism. Jung and Swinney \cite{beck.jung-swinney} have also
experimentally confirmed a simple scaling relation between
$\beta$ and the fluctuating energy dissipation $\epsilon$.

\begin{figure}
\centerline{\epsfxsize=0.6\textwidth\epsffile{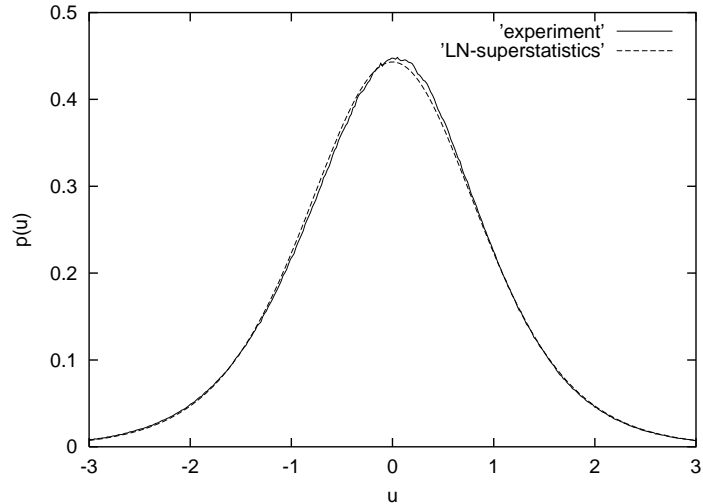}} 
\caption{Experimentally
measured histogram of velocity differences in a Taylor-Couette
experiment \cite{beck.BLS}, and comparison with a superstatistical prediction
($s^2=0.28$).}
\end{figure}

It should be noted that if we know the probability densities
$p(u)$ analytically, as well as the dependence of the parameter $s^2$
on the scale $r$, we can also calculate moments of velocity
differences and thus determine scaling exponents $\zeta_m$ defined by

\begin{displaymath}
\langle u^m \rangle \sim r^{\zeta_m}.
\end{displaymath}
Many different models of $\zeta_m$ can be constructed in such a
way \cite{beck.beck03, beck.beck-sca, beck.chaso}. For further
stochastic models, see e.g.\
\cite{beck.aringazin, beck.dubrulle, beck.greiner}.


\bbib
\bibitem{beck.boden1} A. La Porta, G.A. Voth, A.M. Crawford, J. Alexander,
and E. Bodenschatz, Nature {\bf 409}, 1017 (2001)
\bibitem{beck.boden2} G.A. Voth et al., J. Fluid Mech. {\bf 469}, 121 (2002)



\bibitem{beck.reynolds-recent} A.M. Reynolds, N. Mordant, A.M. Crawford
and E. Bodenschatz, New Journal of Physics {\bf 7}, 58 (2005)
\bibitem{beck.pinton1} N. Mordant, P. Metz, O. Michel, and J.-F.
Pinton, Phys. Rev. Lett. {\bf 87}, 214501 (2001)
\bibitem{beck.pinton2} L. Chevillard,
S.G. Roux, E. Leveque, N. Mordant, J.-F. Pinton, and A. Arneodo,
Phys. Rev. Lett. {\bf 91}, 214502 (2003)
\bibitem{beck.BLS} C. Beck, G.S. Lewis and H.L. Swinney, Phys. Rev. {\bf 63E},
 035303(R) (2001)
\bibitem{beck.beck-physica-d}  C. Beck,
Physica {\bf 193D}, 195 (2004)
\bibitem{beck.jung-swinney} S. Jung and H.L. Swinney, cond-mat/0502301
\bibitem{beck.rapisarda} S. Rizzo and A. Rapisarda,
in Proceedings of the 8th Experimental Chaos Conference, Florence,
AIP Conf. Proc. {\bf 742}, 176 (2004) (cond-mat/0406684)
\bibitem{beck.rap2} S. Rizzo and A. Rapisarda, cond-mat/0502305
\bibitem{beck.peinke} F. B\"ottcher, St. Barth, and J. Peinke,
nlin.AO/0408005
\bibitem{beck.daniels} K.~E. Daniels, C. Beck, and E. Bodenschatz,
Physica {\bf 193D}, 208 (2004)
\bibitem{beck.beck-cohen}
C. Beck and E.G.D. Cohen, Physica {\bf 322A}, 267 (2003)
\bibitem{beck.prl} C. Beck, Phys. Rev. Lett. {\bf 87}, 180601 (2001)
\bibitem{beck.reynolds} A. Reynolds, Phys. Rev. Lett. {\bf 91}, 084503 (2003)
\bibitem{beck.beck03} C. Beck, Europhys. Lett. {\bf 64}, 151 (2003)
\bibitem{beck.castaing} B. Castaing, Y. Gagne, and E.J. Hopfinger, Physica {\bf 46D},
177 (1990)







\bibitem{beck.touchbeck} H. Touchette and C. Beck, Phys. Rev. {\bf 71E}, 016131
(2005)
\bibitem{beck.sawford} B.L. Sawford, Phys. Fluids {\bf A3}, 1577 (1991)

\bibitem{beck.pope} S.B. Pope, Phys. Fluids {\bf 14}, 2360 (2002)
\bibitem{beck.tsa1} C. Tsallis, J. Stat. Phys. {\bf 52}, 479 (1988)

\bibitem{beck.k62} A.N. Kolmogorov, J. Fluid Mech. {\bf 13}, 82 (1962)
\bibitem{beck.beck-sca} C. Beck, Physica {\bf 295A}, 195 (2001)
\bibitem{beck.chaso} C. Beck, Chaos, Solitons and Fractals {\bf 13},
499 (2002)
\bibitem{beck.aringazin} A.K. Aringazin and M.I. Mazhitov,
cond-mat/0408018
\bibitem{beck.dubrulle} J.-P. Laval, B. Dubrulle, and S. Nazarenko,
Phys. Fluids {\bf 13}, 1995 (2001) 
\bibitem{beck.greiner} B. Jouault, M. Greiner, P. Lipa , Physica {\bf 136D},
125 (2000)
\ebib
\end{document}